\documentclass[12pt]{article}
\usepackage{amsfonts}
\usepackage{amsmath}
\usepackage{amssymb}
\usepackage{graphicx}
\usepackage{color}
\usepackage[all, knot]{xy}
\usepackage{tikz}

\usepackage{ulem}

\usepackage[utf8]{inputenc}
\usepackage{epstopdf}
\usepackage[footnotesize]{caption}
\usepackage{amsthm}
\usepackage{enumitem}
\usepackage{mathrsfs}

\usepackage[margin=3cm]{geometry}

\def \be {\begin{equation}}
\def \ee {\end{equation}}
\def \bea {\begin{eqnarray}}
\def \eea {\end{eqnarray}}
\def \nn {\nonumber}

\def \rr {\raise.35ex\hbox{\small $\prime$}\kern-.17em{\mbox{\large $\imath$}}}

\def \dels {\partial\kern-.6em /\kern.1em}
\def \As {{A\kern-.5em / \kern.5em}}
\def \Ds {D\kern-.7em / \kern.5em}

\def \ks {k\kern-.5em /}
\def \ls {l\kern-.5em /}

\newcommand{\ci}[1]{}

\newcommand{\ba}{\begin{eqnarray}}
\newcommand{\ea}{\end{eqnarray}}
\newcommand{\bal}{\begin{align}}
\newcommand{\eal}{\end{align}}
\newcommand{\bay}[1]{\left(\begin{array}{#1}}
\newcommand{\eay}{\end{array}\right)}

\newcommand{\hide}[1]{}

\newlist{axioms}{enumerate}{2}
\setlist[axioms,1]{label=\textbf{A\arabic{axiomsi}.}, ref=A\arabic{axiomsi}}
\setlist[axioms,2]{label=\textbf{A\arabic{axiomsi}\rlap{\myEnumCounter{axiomsii}}.},%
                   ref=A\arabic{axiomsi}\myEnumCounter{axiomsii},%
                   align=parleft,%
                   leftmargin=0em,%
                   itemsep=1.4ex,%
                   before={\stepcounter{axiomsi}}}

  \usetikzlibrary{decorations.markings}

\begin{document}

\begin{titlepage}
\begin{center}

\textbf{\LARGE
More Stringy Effects in Target Space from 
Double Field Theory  
\vskip.3cm
}
\vskip .5in
{\large
Chen-Te Ma$^{a, b, c, d}$ \footnote{e-mail address: yefgst@gmail.com} and Franco Pezzella$^e$ \footnote{e-mail address: franco.pezzella@na.infn.it}
\\
\vskip 1mm
}
{\sl
$^a$
Guangdong Provincial Key Laboratory of Nuclear Science,\\ 
Institute of Quantum Matter,
South China Normal University, Guangzhou 510006, Guangdong, China
\\
$^b$
School of Physics and Telecommunication Engineering,\\
 South China Normal University, 
 Guangzhou 510006, Guangdong, China.
\\
$^c$
The Laboratory for Quantum Gravity and Strings,\\
 Department of Mathematics and Applied Mathematics,
University of Cape Town, Private Bag, Rondebosch 7700, South Africa.
\\
$^d$
Department of Physics and Center for Theoretical Sciences,\\
 National Taiwan University,
 Taipei 10617, Taiwan, R.O.C..
\\
$^e$
Istituto Nazionale di Fisica Nucleare - Sezione di Napoli,\\
Complesso Universitario di Monte S. Angelo ed. 6,\\
 via Cintia,  80126 Napoli, Italy.
}
\\
\vskip 1mm
\vspace{30pt}
\end{center}
\newpage
\begin{abstract}
In Double Field Theory, the mass-squared of doubled fields associated with bosonic closed string states is proportional to $N_L+N_R-2$. 
Massless states are therefore not only the graviton, anti-symmetric, and dilaton fields with $(N_L=1, N_R=1)$ such theory is focused on, but also the symmetric traceless tensor and the vector field relative to the states $(N_L=2, N_R=0)$ and $(N_L=0, N_R=2)$ which are massive in the lower-dimensional non-compactified space. 
While they are not even physical in the absence of compact dimensions, they provide a sample of states for which both momenta and winding numbers are non-vanishing, differently from the states $(N_L=1, N_R=1)$. 
A quadratic action is therefore here built for the corresponding doubled fields.
It results that its gauge invariance under the linearized double diffeomorphisms is based on a generalization of the usual weak constraint, giving rise to an extra mass term for the symmetric traceless tensor field, not otherwise detectable: this can be interpreted as a mere stringy effect in target space due to the simultaneous presence of momenta and windings.
Furthermore, in the context of the generalized metric formulation, a non-linear extension of the gauge transformations is defined involving the constraint extended from the weak constraint that can be uniquely defined in triple products of fields.  
Finally, we show that the above mentioned stringy effect does not appear in the case of only one compact doubled space dimension.
\end{abstract}
\end{titlepage}

\section{Introduction}
\label{sec:1}
\noindent
When compactified on a $d$-dimensional torus $T^{d}$, string theory exhibits the peculiar symmetry O($d$, $d$; $\mathbb{Z}$) \cite{Maharana:1992my} for all the $d$ compact directions \cite{Giveon:1991jj}: the target-space duality (T-duality) \cite{Buscher:1987sk, Buscher:1987qj}. 
It is a distinctive symmetry of strings since, differently from particles, one-dimensional objects can wrap $d$ non-contractible cycles. 
A winding number counts the times a string wraps around a circle in the target space. 
Winding modes $\omega^{a}\equiv m_{a} R_a/\alpha^{\prime}$ $(a=1, 2, \cdots, d;\ m_{a}\in\mathbf{Z})$ around the compact circle of radius $R_{a}$ of $T^{d}$ with coordinate $x^{a}$ have to be added to Kaluza-Klein momentum modes $p_{a}=n_{a}/R_{a}$. 
The O($d$, $d$; ${\mathbb Z}$) T-duality is a symmetry that establishes a connection between the two different but dual tori: $T^{d} $ and $\tilde{T}^{d}$. 
For a rectangular torus, a T-duality transformation consists in exchanging momentum and winding modes while mapping the circle of radius $R_{a}$ of $T^{d}$ with coordinate $x^{a}$ into the dual circle, with coordinate $\tilde{x}_{a}$ and periodicity $2 \pi \alpha^{\prime} / R_{a}$, where $\alpha^{\prime}$ is the Regge slope parameter. 
While the momentum $p_{a}$ is the conjugate variable to $x^{a}$, the winding mode results to be the conjugate variable of the coordinate $\tilde{x}_{a}$. 
The most intuitive realization of a T-dual invariant formulation of string theory is to introduce a manifest symmetry between windings $\omega^{a}$ and Kaluza-Klein momenta $p_{a}$ \cite{Siegel:1993xq, Siegel:1993th} or, equivalently, between $x^{a}$ and $\tilde{x}_{a}$ and hence between the string coordinates $X^{a}$  and their duals $\tilde{X}_{a}$ already at the level of the string world-sheet. This should generate a manifestly T-dual invariant formulation of the corresponding target space theory \cite{Hitchin:2004ut, Gualtieri:2003dx, Duff:1989tf, Hohm:2010pp, Tseytlin:1990nb, Ma:2017jeq}. 
Having a T-dual invariant formulation of string theory possibly also has the advantage of providing a field theory description of winding states, not reachable through the usual field theory limit $\alpha^{\prime} \rightarrow 0$. 
\\

\noindent
The above goal can be pursued  both within the first-quantized string theory and in the context of the second-quantized string theory, in particular of the closed string field theory \cite{Zwiebach:1992ie, Hata:1986mz, Kugo:1992md, Alvarez:1996vt, Ghoshal:1991pu}.   
The closed string field theory on a $T^{d}$ torus is naturally formulated in such a way that it exhibits a manifest invariant T-dual structure since string fields are necessarily defined on the $2d$-dimensional doubled torus, formed by the coordinates $x^{a}$ of $T^{d}$ and the coordinates $\tilde{x}_a$ of the dual torus $\tilde{T}^{d}$. 
Inspired by these features of closed string field theory, Double Field Theory \cite{Hull:2009mi} is a proposal to incorporate T-duality  already as a symmetry structure of a field theory. 
Geometry underlying Double Field Theory is novel and when restricted to a half-dimensional space, it includes Generalized Geometry, based on substituting the tangent space in each point of the target space with a direct sum of the tangent and the cotangent spaces \cite{Hitchin:2004ut, Gualtieri:2003dx, Duff:1989tf, Hohm:2010pp, Tseytlin:1990nb, Ma:2017jeq}.
\\

\noindent
The fields $\phi_{I}(x^{\mu}, x^{a}, \tilde{x}_{a})$ of Double Field Theory remember the constraints imposed on the corresponding physical string states. 
On-shell string physical states need to be annihilated by the level matching condition $L_{0} - \bar{L}_{0}=0$ and by the free string on-shell condition, where $L_{0}$ and $\bar{L}_{0}$ are the well-known Virasoro operators, defined in terms, respectively, of the string left and right modes, respectively. 
The former gives rise to the condition $N_L-N_R-\alpha^{\prime}p_a w^a=0$ with $N_L$ and $N_R$ being the number of left-moving and right-moving oscillators, while the latter allows one to determine the squared mass of the corresponding physical state. 
The definition of the squared mass
\bea
 M^{2} \equiv - (k^{2} + p^{2} + \omega^{2})  \label{squaredm}
\eea 
of a physical string state in $all$ of the $D$ (non-compact and compact) dimensions of the target space involves symmetrically the momenta along the non-compact directions $k_{\mu}$, the momenta along the compact directions $p_{a}$, and the winding $\omega^{a}$ with $ p^{2} = \hat{G}^{ab} p_{a}p_{b}$ and $ \omega^{2}= \hat{G}_{ab} \omega^{a} \omega^{b}$, being $\hat{G}_{ab}$ the torus metric given by $\hat{G}_{ab} =\delta_{ab} R^{2}_{a}/\alpha^{\prime}$ for a rectangular torus. 
A simple expression for $M^{2}$ is obtained when the background Kalb-Ramond field vanishes
\bea
M^{2}= \frac{2}{\alpha^{\prime}}(N_L+N_R-2).   \label{massD-dim}
\eea
For $N_L=N_R=1$, Eq. \eqref{massD-dim} defines a set of massless fields living in $D$ dimensions that would also be massless also in the non-compactified theory. 
They have the same index structure as in the non-compact directions but keep their full dependence both on the coordinates of the doubled torus and the non-compact ones. 
These fields result to be: $h_{jk} (x^{\mu}, x^{a}, \tilde{x}_{a})$, $b_{jk} (x^{\mu}, x^{a}, \tilde{x}_{a})$, and $\phi(x^{\mu}, x^{a}, \tilde{x}_{a})$ with  $\,\,\,j, k=1, 2,\cdots, D \,\,\,$; $\mu, \nu = 0, 1, \cdots, D-d$, and $a=1, 2, \cdots, d$  and these are the fields  on which Double Field Theory is constructed \cite{Hull:2009mi}. 
It turns out that the so-called weak constraint $\partial_a\partial^{\tilde{a}}f =0$ has to be imposed on them in order to have a consistently gauge-invariant theory under diffeomorphisms and anti-symmetric tensor gauge transformations \cite{Hull:2009mi}. 
When imposing the weak constraint, the above fields $h_{jk}, b_{jk}$ and $\phi$ depend only on $(x^{\mu}, x^{a})$ or, alternatively, on $(x^{\mu}, \tilde{x}_{a})$ providing respectively the familiar tensor metric, the Kalb-Ramond and the dilaton in $D$ dimensions or their dual versions. 
The weak constraint $\partial_a\partial^{\tilde{a}} f=0$ is reminiscent of the level matching condition $p_{a} \omega^{a} = 0$ applied in the particular case of $N_{L}=N_{R}=1$, being $\partial_a\equiv \partial/\partial x^a$ and $\tilde{\partial}^a\equiv\partial/\partial \tilde{x}_a$ the operators respectively associated with $p_{a}$ and $\omega^{a}$. 
It is clear from its definition that the weak constraint eliminates the possibility of having doubled fields with a dependence on both momenta and windings, as simply dictated by the level matching condition. 
Furthermore, the weak constraint applied to a product of fields gives rise to the definition of the strong constraint, introduced for having a manifestly O($D$, $D$) structure in target space \cite{Hohm:2010pp}, where an extension of T-duality is realized from O($d$, $d$; $\mathbb{Z}$) to O($D$, $D$) by associating with the non-compact dimensions $x^{\mu}$ and the corresponding null dual coordinates $\tilde{x}_{\mu}$. 
This allows one to treat non-compact and compact dimensions in a symmetric way.  
The strong constraint is necessary for non-compact directions from the string perspective \cite{Betz:2014aia, Aldazabal:2011nj, Geissbuhler:2011mx}.
\\
 
 \noindent
 Let us remind here, as already stressed in Ref. \cite{Hull:2009mi}, that the definition of squared mass in Eq. \eqref{squaredm} is different from the corresponding one given by:  
 \bea
 {\cal M}^{2} \equiv  - k^{2} = p^{2} + \omega^{2} +\frac{2}{\alpha^{\prime}}(N_L+N_R-2) =   p^{2} + \omega^{2} + M^{2} 
 \label{m2lowdim}
 \eea
in the non-compact $(D-d)$-dimensional Minkowski space where, therefore, a conventional effective theory would keep states with zero or small values of ${\cal M}^{2}$. 
Hence the spectrum of the $D$-dimensional states with $M^{2}=0$ does not coincide with the analogous massless spectrum of particles in the lower $(D-d)$-dimensional theory and, in particular, does not include the $d$-dimensional extra vector states $(N_L=1, N_R=0)$ and $(N_L=0, N_R=1)$ with ${\cal M}^2=0$ giving an enhanced gauge symmetry at the self-dual compactification radius $R_{a} = \sqrt{\alpha^{\prime}}$ \cite{Aldazabal:2015yna, Aldazabal:2018uzm, Fraiman:2018ebo}. 
Instead, Double Field Theory keeps states with $M^2=0$ that include, therefore, not only the states $(N_L=1,N_R=1)$, already mentioned above but also the states $(N_L=2, N_{R}=0)$ and $(N_L=0, N_{R}=2)$. 
They have vanishing squared mass $M^{2}$ in $D$ dimensions according to Eq. \eqref{massD-dim} but correspond to massive states in $(D-d)$ dimensions with ${\cal M}^2$. 
There are the states on which this work is focused on. 
The reason why they are interesting is that the level matching condition applied to such states, $\alpha^{\prime}p_a w^a=2$ and $\alpha^{\prime}p_a w^a=-2$, implies for them a simultaneous presence of non-vanishing momentum and winding modes, differently from the more familiar massless state $(N_L=1, N_R=1)$ for which, instead, such simultaneous presence is inhibited.  
Constructing, through Double Field Theory, a theory of doubled fields corresponding to such string states could reveal therefore field theoretical aspects due to the simultaneous presence of momenta and windings and could shed light on more stringy effects in target space which would be difficult to capture otherwise. 
The action will provide an answer to the central question addressed in this work: {\it What is the target space theory that highlights more stringy features in Double Field Theory?} 
In other words, what is the target space field theory that could incorporate the simultaneous presence of  momenta and winding modes allowed by a deformation of the weak constraint \cite{Ma:2016vgq}? 
This investigation is a first step to extending Double Field Theory beyond the supergravity spectrum.
 
\section{Quadratic Theory}
\label{sec:2}
\noindent
In this section, we strictly follow Ref. \cite{Hull:2009mi}, and the quadratic action for the fields corresponding to the string states $(N_L=1, N_R=1)$ will be here borrowed in order to write the analogous action for the fields corresponding to the string states $(N_L=2, N_R=0)$ and $(N_L=0, N_R=2)$ also with $M^{2}=0$, provided that a suitable correspondence can be established between the two families of states in the two cases.
\\

\noindent
Let us remind that, for the case $(N_L, N_R)=(1, 1)$, the quadratic action in Double Field Theory is \cite{Hull:2009mi}
\bea
&&S^{(2)}_{DFT}
\nn\\
&=&\frac{1}{16\pi G_{N}}\int [dxd\tilde{x}] \ \bigg(\frac{1}{4}h^{jk}\partial_{l}\partial^{l}h_{jk}
+\frac{1}{2}\partial^{j}h_{jk}\partial_{l}h^{lk}
-2 \Phi \partial^{j}\partial^{k}h_{jk}-4 \Phi \partial^{j}\partial_{j} \Phi
\nn\\
&&+\frac{1}{4}h^{jk}\tilde{\partial}_{l}\tilde{\partial}^{k}h_{jk}
+\frac{1}{2}\tilde{\partial}^{j}h_{jl}\tilde{\partial}_{k}h^{k l}
+2 \Phi  \tilde{\partial}^{j}\tilde{\partial}^{k}h_{jk}-4 d \tilde{\partial}^{j}\tilde{\partial}_{j} \Phi
\nn\\
&&+\frac{1}{4}b^{jk}\partial^{l}\partial_{l}b_{jk}
+\frac{1}{2}\partial^{k}b_{jk}\partial_{l}b^{j l}
+\frac{1}{4}b^{jk}\tilde{\partial}^{l}\tilde{\partial}_{l}b_{jk}
+\frac{1}{2}\tilde{\partial}^{k}b_{jk}\tilde{\partial}_{l}b^{jl}
\nn\\
&&+(\partial_{k}h^{j k})\tilde{\partial}^{l}b_{jl}
+(\tilde{\partial}^{l}h_{i k })\partial_{l}b^{jk}
-4 \Phi \partial^{i}\tilde{\partial}^{j}b_{ij}\bigg),   \label{s11}
\eea
where $G_{N}$ is the gravitational constant, and $\int [dxd\tilde{x}]$ is an integral over all of the $n+2d$ coordinates of $\mathbb{R}^{n-1,1} \times T^{2d}$ being this latter the doubled torus with periodic coordinates $(x^{a}, \tilde{x}_{a})$.
\\

\noindent
The action $S^{(2)}_{DFT}$ in Eq. \eqref{s11} describes the dynamics of the fluctuations $h_{jk}(x^{\mu}, x^{a}, \tilde{x}_{a})$ and $b_{jk}(x^{\mu}, x^{a}, \tilde{x}_{a})$ around constant backgrounds  $G_{jk}$ and $B_{jk}$ respectively. 
Indices are raised and lowered by $G_{jk}$. 
Furthermore, the constant toroidal background field $E_{jk}=G_{jk} + B_{jk}$ and, correspondingly, the fluctuations $e_{jk} = h_{jk} + b_{jk}$ can be introduced. For simplicity, backgrounds with $B_{jk}=0$ are considered. 
The field $\Phi(x^{\mu}, x^{a}, \tilde{x}_{a})$ corresponds to the scalar dilaton, invariant under T-duality with its expectation value providing the duality invariant string coupling constant. 
\\

\noindent
The gauge invariance of $S^{(2)}_{DFT}$ is respect to linear doubled diffeomorphisms generated by the vector fields $\epsilon_{j} (x^{\mu}, x^{a}, \tilde{x}_{a})$ and $\tilde{\epsilon}_{j}(x^{\mu}, x^{a}, \tilde{x}_{a})$ given by:
\bea
\delta h_{jk} & = & \partial_{j}\epsilon_{k}+\partial_{k}\epsilon_{j}
+\tilde{\partial}_{j}\tilde{\epsilon}_{k}+\tilde{\partial}_{k}\tilde{\epsilon}_{j} \, ; 
\nn\\
\delta b_{jk} & = & -\big(\tilde{\partial}_{j}\epsilon_{k}-\tilde{\partial}_{k}\epsilon_{j}\big) 
-\big(\partial_{j}\tilde{\epsilon}_{k}-\partial_{k}\tilde{\epsilon}_{j}\big) \,;  \label{gaugetransf}
\nn\\
\delta \Phi & = & -\frac{1}{2} ( \partial_{j}\epsilon^{j}-\tilde{\partial}_{j}\tilde{\epsilon}^{j}) \, .
\eea
When $\epsilon_{j}$ and the fields themselves are independent of $\tilde{x}_{j} \equiv (\tilde{x}_{a}, \tilde{x}^{\mu}=0)$, then the  above transformations reproduce the standard linearized diffeomorphisms with parameter $\epsilon_{j}$ involving the coordinates $x^{j}$ and the anti-symmetric tensor gauge transformations with parameters $\tilde{\epsilon}_{j}$. 
Analogously, fields and parameters independent of $x^{i} \equiv (x^{\mu}, x^{a})$ are defined on the dual space with the roles of the parameters $\epsilon_{j}$ and $\tilde{\epsilon}_{j}$, interchanged in the doubled diffeomorphisms. 
In fact, diffeomorphisms and anti-symmetric gauge transformations are strictly linked, and their variables are interchanged by T-duality. The T-duality invariance means that the action $S^{(2)}_{DFT}$ remains unchanged under an O($d$, $d$; $\mathbb{Z}$), i.e. a $2d \times 2d$ transformation matrix $g$ relating $E_{jk}$ and $E^{\prime}_{jk}$ as follows:
\bea
E^{\prime}= g(E) = \frac{a E + b}{cE + d};\ 
g = 
\begin{pmatrix}
a& b
\\
c& d
\end{pmatrix};\ 
\eta^Tg\eta=\eta;\ 
\eta = 
\begin{pmatrix}
0& \mathbf{1} 
\\ 
\mathbf{1} & 0 
\end{pmatrix}
\eea
with $a$, $b$, $c$, and $d$ being $d\times d$ matrices, and $\eta$ being the O($d$, $d$) invariant metric. 
\\

\noindent
Furthermore, as already observed above, while the scalar dilaton $\Phi$ is invariant under T-duality, there is instead no dilaton that is a scalar under both diffeomorphisms and dual diffeomorphisms. 
One can actually define a dilaton $\phi\equiv  \Phi + G^{jk}h_{jk}/4$, invariant under the standard linearized transformations acting on $x^{j} = (x^{\mu}, x^{a})$. 
An analogous definition can be given for a dual dilaton $\tilde{\phi }\equiv  \Phi - G^{jk}h_{jk}/4$ under the dual diffeomorphisms generated by $\tilde{\epsilon}_{j}$ and acting on $\tilde{x}_{j}$. 
Non-linearly, a relation of the form $\exp(-2\Phi)\equiv \exp(-2\phi)\sqrt{-\det h_{jk}}$ holds \cite{Hull:2009mi}. 
It has to be stressed that the gauge invariance of $S^{(2)}_{DFT}$ holds only if the weak constraint $\partial_j\tilde{\partial}^jf=0$ is imposed on fields and gauge parameters \cite{Hull:2009mi}.
\\

 \noindent
 The action \eqref{s11} is the starting point for the construction of the quadratic Double Field Theory action for the fields corresponding to the string states with $(N_L=2, N_R=0)$ and $(N_L=0, N_R=2)$. 
 Before doing that, let us first discuss the content of these levels in string theory. 
 We shall consider the critical dimensionality $D=26$ of bosonic closed string theory, but the results will be also valid for closed superstring theories. 
 For illustrative purposes, we consider the case of one coordinate compactified on a circle of radius $R$, e.g. $X^{25}$ satisfying the periodicity condition $X^{25} \sim X^{25} + 2 \pi R m$ with $m \in \mathbb{Z}$. 
 Physical states have to satisfy the constraint $N_{L} - N_{R} = \alpha^{\prime} \, {p}_{25} \,\, \omega^{25}=nm$ with $p_{25}=n/R$ and $\omega^{25} = mR/\alpha^{\prime}$. 
 This means that for the states $(N_L=2, N_R=0)$ one has $(n, m)=(1, 2), (2, 1), (-1, -2), (-2, -1)$ while the states $(N_L=2, N_R=0)$ require $(n, m)=(-1, 2), (-2, 1), (1, -2), (2, -1)$. 
  \\
  
  \noindent
  For each of these possibilities, the physical state content for the level $(N_L=2, N_R=0)$ $[\mbox{resp.} \,\,(N_L=0, N_R=2)]$ is generated by the action of the light-cone left $[\mbox{resp.} \,\mbox{right}]$ creation moving oscillators $\big( \alpha_{-1}^{j} \big) \big(\alpha_{-1}^{k}\big)$ or $\big(\alpha_{-2}^{j}\big)$ [resp. $\big(\bar{\alpha}_{-1}^{j}\big) \big(\bar{\alpha}_{-1}^{k}\big)$ or $\big(\bar{\alpha} _{-2}^{j }\big)$] on the vacuum tachyon state. 
  In such a case, the product of creation operators $\big(\alpha_{-1}^{j}\big) \big(\alpha_{-1}^{k}\big)$ generates a symmetric traceless tensor with $(D-2)(D-1)/2-1$ physical degrees of freedom and a scalar with one degree of freedom. 
  We denote the doubled fields associated with them again respectively by $h_{jk}(x^{\mu}, x^{a}, \tilde{x}_{a})$ and $\Phi (x^{\mu}, x^{a}, \tilde{x}_{a})$. 
 The second creation operator $\big(\alpha_{-2}^{j}\big)$ defines a vector state and its dual, described by a one-form gauge field $A_j(x^{\mu}, x^{a}, \tilde{x}_{a})$ with $(D-2)$ physical degrees of freedom. 
 \\
 
 \noindent 
 All these states are massless in $D$-dimensions, with $M^{2}=0$, but they are massive in the lower-dimensional $(D-d)$-dimensional non-compact spacetime, where Kaluza-Klein momenta and windings contribute to the squared-mass according to ${\cal M}^2$ in Eq. \eqref{m2lowdim}. 
\\

\noindent
Summarizing, the fields associated with the string states of the levels $(N_L=2, N_R=0)$ and $(N_L=0, N_R=2)$ correspond to a symmetric traceless tensor $h_{jk}(x^{\mu}, x^{a}, \tilde{x}_{a})$, a one-form gauge field $A_j(x^{\mu}, x^{a}, \tilde{x}_{a})$,  and a scalar field $\Phi (x^{\mu}, x^{a}, \tilde{x}_{a})$ with a missing anti-symmetric field. 
Actually, the one-form gauge field $A_j$ can be used to define still an anti-symmetric tensor
\bea
b_{jk} =  -\big(\tilde{\partial}_{j}A_{k}-\tilde{\partial}_{k}A_{j}\big)
+\big(\partial_{k}A_{j}-\partial_{j}A_{k}\big) \,\, .   \label{bjk}
\eea
The gauge transformation of $A_{j}$ is provided by $\delta A_{j}$ such that 
\bea
\delta A_{j}  =   \epsilon_{j} = \tilde{\epsilon}_{j},
\eea
 as suggested by the gauge transformation of $b_{jk}$ in Eq. \eqref{gaugetransf}. 
 This implies $\epsilon_{j} = \tilde{\epsilon}_{j}$, and we shall see in a while that this identification will play a relevant role. 
\\

\noindent
In conclusion, it results that the fields of the level  $(N_L=2, N_R=0)$ and $(N_L=0, N_R=2)$ can be put in a one-to-one correspondence with the ones of the level $(N_L=1, N_R=1) $ \cite{Hull:2009mi}. 
This correspondence allows us to consider the quadratic action in Eq. \eqref{s11} as our starting point, but adapted to the double states $(N_L=2, N_R=0)$ or $(N_L=0, N_R=2)$. 
We will denote by $\tilde{S}_{new}^{(2)}$ the quadratic action after applying the above-mentioned correspondence of fields, with the aim of distinguishing the two cases $S^{(2)}_{DFT} \rightarrow \tilde{S}^{(2)}_{new}$.
\\

\noindent
The variation $\delta S^{(2)}_{DFT}$ of the action \eqref{s11} with respect to the gauge transformations in Eq. \eqref{gaugetransf} vanishes only if the weak constraint $\partial_{a} \tilde{\partial}^{a}f=0$ is imposed on the gauge parameters and double fields. 
Therefore, in the cases $(N_L=2, N_R=0)$ and $(N_L=0, N_R=2)$, we  start from $\tilde{S}^{(2)}_{new}$, with the only modification to do concerning the weak constraint that now becomes \cite{Hull:2009mi}:
\bea
 \partial^{a} \tilde{\partial}_{a} f= -\frac{(N_L-N_R)}{\alpha^{\prime}}f\equiv-\frac{\lambda}{2} f.
 \label{newconstr}
\eea
As already observed, also this constraint is reminiscent of the level matching condition $N_{L} - N_{R} - \alpha^{\prime} p_{a} w^{a}=0$, but this time applied to the case $N_{L} \neq N_{R}$. 
It is straightforward to calculate the variation of $\tilde{S}^{(2)}_{new}$ under the same linear gauge transformations listed in Eq. \eqref{gaugetransf}. 
While $ S^{(2)}_{DFT}$  is invariant under those gauge transformations when the gauge parameters are subject to the weak condition $\partial_{a} \tilde{\partial}^{a} f = 0$, the analogous variation of $\bar{S}^{(2)}_{new}$ is not zero and results to be
\bea
\delta  \tilde{S}^{(2)}_{new} = \frac{1}{16\pi G}\int dxd\tilde{x}\ \bigg(\frac{\lambda}{4}\delta(b^{jk}b_{jk})+\lambda h^{jk}(\partial_{k}\tilde{\epsilon}_{j}+\tilde{\partial}_{k}\epsilon_{j})+4\lambda \Phi(\partial^{k}\tilde{\epsilon}_{k}-\tilde{\partial}^{k}\epsilon_{k})\bigg)
\nn\\
\eea
with fields and gauge parameters now subject to the new modified constraint \eqref{newconstr}.
Let us assume $\epsilon_{j}=\tilde{\epsilon}_{j}$, as already requested by the variation of the anti-symmetric field $b_{jk}$ from Eq. \eqref{bjk}. 
One obtains
\bea
\delta \tilde{S}^{(2)}_{new} 
=\frac{1}{16\pi G_N}\int dxd\tilde{x}\ \bigg(\frac{\lambda}{4}\delta(b^{jk}b_{jk})+\frac{\lambda}{4}\delta(h^{jk}h_{jk})-4 \lambda \delta (\Phi^{2})\bigg).
\eea
In order to have a quadratic action invariant under the generalized transformations, the following term $\tilde{S}^{(2)}_{add} $ has therefore to be added to $\bar{S}^{(2)}_{new} $ in order to cancel the non-invariant terms
\bea
\tilde{S}^{(2)}_{add}=\frac{1}{16\pi G_N}\int dxd\tilde{x}\ \bigg(-\frac{\lambda}{4}b^{jk}b_{jk}-\frac{\lambda}{4}h^{jk}h_{jk}+4\lambda \Phi^2\bigg).
\eea
The quadratic action of Double Field Theory $\tilde{S}^{(2)}_{DFT}$ for the states $(N_{L}=2, N_{R}=0)$ and $(N_L=0, N_R=2)$ is therefore given by:
\bea
\tilde{S}^{(2)}_{DFT}  & =  &  \tilde{S}^{(2)}_{new} +  \tilde{S}^{(2)}_{add}  \nonumber \\ 
{} & =  & \tilde{S}^{(2)}_{new} + \frac{1}{16\pi G_N}\int dxd\tilde{x}\ \bigg(-\frac{\lambda}{4}b^{jk}b_{jk}-\frac{\lambda}{4}h^{jk}h_{jk}+4\lambda \Phi^2\bigg).
\eea
\\

\noindent
This shows that the parameter $\lambda$ appearing in the new constraint \eqref{newconstr} gives a mass term to the fluctuation fields $h_{jk}, b_{jk}$ and the scalar dilaton field $\Phi$. 
Furthermore, the identification $\epsilon_{j}=\tilde{\epsilon}_{j}$ that makes the theory gauge invariant under the generalized diffeomorphisms by using the new constraint, now creates an interdependence of the fields in the target space and the dual fields in the dual target space: they, therefore, do not constitute independent physical degrees of freedom.
\\
   
\noindent
When the dilaton field $\phi$ vanishes, the scalar dilaton field $\Phi$ at the quadratic order is $ \Phi=-h^{j}{}_{j}/4$. 
Consequently, a graviton-like massive term $\lambda\big(h^{jk}h_{jk}-(h^j{}_j)^2\big)$ is generated for the symmetric traceless field showing that the stringy effect is mainly embodied in the parameter $\lambda$. 
This provides a graviton-like massive term in the theory of the target space that, from the point of view of the non-compact lower $(D-d)$-dimensional spacetime, is given by ${\cal M}^{2}_{g} =  p^{2} + \omega^{2}    +  2(N_L-N_R)/\alpha^{\prime}$.
Furthermore, it is worth observing that the appearance of the new gravitational mass term is uniquely due to the non-simultaneous vanishing of momenta and windings, which is provided by the states $(N_{L},N_{R}) = (2,0),(0,2)$ in consideration. 
It could be therefore interpreted as due to an interaction between momentum and winding modes. 
\\

\noindent
Finally, we give some comments on the T-duality \cite{Buscher:1987sk, Buscher:1987qj} for the quadratic action $\tilde{S}^{(2)}_{DFT}$. 
As already mentioned, the latter, just like $S_{DFT}^{(2)}$, represents the dynamics of the fluctuations $h_{jk}$ and $b_{jk}$ around a suitable background $E_{jk}$ and scalar dilaton $\Phi$. 
The T-duality invariance of the action $S_{DFT}^{(2)}$ can be extended to the action $\tilde{S}_{DFT}^{(2)}$ without any role played by the modified weak constraint, meaning that the T-duality invariance, at the quadratic level, does not see the deformation carried out by $\lambda$ in the constraint \eqref{newconstr}.
\\

\noindent
In Ref. \cite{Hull:2009mi}, beyond the quadratic action in the fluctuation fields described by the action $S^{(2)}_{DFT}$, also a cubic action with the corresponding gauge transformations have been constructed by circumventing a problem related to non-associativity. 
The resulting theory does not exhibit a manifest O($d$, $d$; $\mathbb{Z}$) symmetry. 
For having a manifest T-dual invariant theory, the first necessary step is to perform an extension of T-duality from O($d$, $d$; $\mathbb{Z}$) to O($D$, $D$). 
This is done by associating the non-compact dimensions $x^{\mu}$ with the corresponding dual coordinates $\tilde{x}_{\mu}=0$ and allowing to treat compact and non-compact dimensions in all the $D$ dimensional target space in a symmetric way through $x^{j}\equiv(x^{\mu}, x^{a})$ and $\tilde{x}_{j} \equiv (\tilde{x}_{\mu}=0, \tilde{x}_{a})$. 
The O($D$, $D$) symmetry is then broken to the subgroup O($d$, $d$; $\mathbb{Z}$) preserving the boundary conditions in the presence of the $d$ compact coordinates. 
An action with a manifest O($D$, $D$) symmetry structure can therefore be obtained by rewriting the action in terms of O($D$, $D$) tensors: the scalar dilaton $\Phi$ and the generalized metric ${\cal H}_{MN}$, defined in terms of the metric tensor field $g$ and the antisymmetric tensor field ${\mbox b}$ as follows
\bea
{\cal H}_{MN}\equiv
\begin{pmatrix}
g^{-1}& -g^{-1}{\mbox b}
\\
{\mbox b}g^{-1}& g-{\mbox b}g^{-1}{\mbox b}  \label{HMN}
\end{pmatrix},
\eea 
where O($D$, $D$) indices $M, N =1, 2, \cdots, 2D$ have been introduced. 
This is the core of the generalized metric formulation of Double Field Theory \cite{Hohm:2010pp}.
\\

\noindent
The weak constraint itself can be rewritten in an O($D$, $D$) covariant form. 
Actually, the two derivatives $\partial_{j}$ with respect to $x^{j}$ and $\tilde{\partial}_{j}$ with respect to $\tilde{x}^{j}$ can be used for defining the partial derivative $\partial_J\equiv
\begin{pmatrix}
\tilde{\partial}^j
&
\partial_{j}
\end{pmatrix}^T$ with an O($D$, $D$) index $J=1, 2, \cdots, 2D$ with the O($D$, $D$) indices being raised or lowered by the 
O($D$, $D$) invariant metric
\bea
\eta_{AB}\equiv
\begin{pmatrix}
0& \bf{1}
\\
\bf{1} &0
\end{pmatrix}.
\eea
This allows to obtain the O($D$, $D$) covariant weak constraint: 
\bea
\partial_{J} \partial^{J}f= 2 \partial_{a} \tilde{\partial}^{a}f = 0
\eea
 in the case $(N_{L}=1, N_{R}=1)$. 
In the cases $(N_L=2, N_R=0)$ and $(N_L=0, N_R=2)$, the modified weak constraint can be written by using O($D$, $D$) indices as $\partial_J\partial^Jf=-\lambda f$: this constraint breaks the O($D$, $D$) structure as the emergence of massive fields witnesses but still one can discuss the gauge transformations for the O($D$, $D$) fields, i.e. the generalized metric and the scalar dilaton, and try to understand how the gauge symmetry could be broken by the constraint.

\section{Gauge Transformation}
\label{sec:3}
\noindent
In any theory with a metric $g_{jk}$ and an anti-symmetric tensor field ${\mbox b}_{jk}$ like in Double Field Theory, diffeomorphisms are generated by vector fields $\xi^{j}$ while anti-symmetric tensor gauge transformations are generated by one-forms $\tilde{\xi}_{j}$. 
This is of course also true in the case of the theory described by the action $\tilde{S}^{(2)}_{DFT}$ for which one can define the gauge transformations generated by the double vector $\xi^{P} = \begin{pmatrix}\tilde{\xi}_{j}  & {\xi}^{j}  \end{pmatrix}^T$ having the two kinds of gauge parameters as components. 
The gauge transformations for $g_{jk}$ and ${\mbox b}_{jk}$ induce a gauge transformation for the generalized metric that, together with the scalar dilaton, is the fundamental field in the generalized metric formulation.
\\

\noindent
Before defining the non-linear gauge transformations of the fields of the theory, we first show how the modified weak constraint could be easily imposed through a suitable $*$ star product operation that is going to be defined in the following. 
It results to be easier to work in the momentum space, in order to project a generic field $A$ down to the physical space with $\partial^{J} \partial_{J} A= - \lambda A$.
\\

\noindent
For a generic double field $A(\tilde{x}_{m},x^{m})$, one can introduce a Fourier series along the dimensions of the doubled space as follows
\bea
A \equiv \sum_{K} A_{K} e^{iKX},
\eea
where $KX \equiv K^{M}X_{M}$. 
Here we define $X^{M} \equiv (\tilde{x}_{j}, x^{j})$ and  $K^{M} \equiv (p_{j}, w^{j})$, with $p_j$ being the momentum along the $j$-th dimension, and $w^j$ being the corresponding winding number with $KX$ defined through the O($D$, $D$) invariant metric $\eta$.
\\

\noindent
The constraint in Eq. \eqref{newconstr} can be rewritten in the O($D$, $D$) notation $\partial_J\partial^JA=-\lambda A$, which can be imposed on the field $A$ by embodying it in the following star product where it is transferred on the momenta as $K_JK^J=\lambda$ \cite{Hull:2009mi} 
\bea
A*1\equiv\sum_K A_K\exp(iKX)\delta_{KK,\lambda} \label{starp}
\nn
\eea
that implies:
\bea
A*B&=&B*A
\nn\\
&\equiv&\sum_{K_A, K_B}A_{K_A}B_{K_B}\exp\big(i(K_A+K_B)X\big)\delta_{K_AK_A, \lambda}\delta_{K_BK_B, \lambda}\delta_{K_AK_B, -\frac{\lambda}{2}}.
\nn\\
\label{AB}
\eea     
In other words,  the star product above defined directly imposes the modified strong constraint Eq. \eqref{newconstr} on $A$ and on the product of constrained fields $A$ and $B$:
\bea
           \partial_J \partial^J (A *1) = - \lambda (A*1)  \,\,\,\,\,\,\,\, \mbox{and}  \,\,\,\,\,\,\,\, \partial_J \partial^J (A*B) = - \lambda (A*B).
           \eea
           \\
           
\noindent
Furthermore, we would like to stress that the star product in Eq. \eqref{starp} also yields the constraint in the triple-products of fields and gauge parameters: $\partial_{M}\partial^{M}\big(({\cal A}*{\cal B})*{\cal C}\big)=-\lambda ({\cal A}*{\cal B})*{\cal C}$. 
The latter condition implies $({\cal A}*\partial^{M}{\cal B})*\partial_{M}{\cal C}=(\lambda/4) ({\cal A}*{\cal B})*{\cal C}$. 
These conditions put in evidence the non-associativity of the above-defined star product $*$ when $\lambda\neq 0$. 
Once again, one can notice that when $\lambda=0$, the constraint is equivalent to the usual strong constraint, but it goes beyond this latter for $\lambda\neq0$.
\\
           
\noindent
The two main fields on which the generalized metric formulation is based are the generalized metric itself and the scalar dilaton, as already claimed. 
These basic fields can be defined, in the theory that we are discussing, in terms of the star product as follows:
\bea
{\cal H}_{MN}\equiv
\begin{pmatrix}
g^{-1}& -g^{-1}*{\mbox b}
\\
{\mbox b}*g^{-1}& \,\,\,g-{\mbox b}*g^{-1}*{\mbox b}   \,\,\,\,\,\,\,   \label{HMN}  
\end{pmatrix};
\,\,\,\,\,\,\,\,\, \Phi = e^{-2 \phi} * \sqrt{| \mbox{det} g |}.
\eea 
\\

\noindent
The above projection realized by the star product $*$ has of course to be used in the non-linear gauge transformations of these two main fields in the terms involving the product of fields and gauge parameters for ensuring that the gauge variations are allowed variations of the fields. 
They are given by:
\bea
\delta_{\xi}{\cal H}^{MN}
&=&\xi^{P}*\partial_{P}{\cal H}^{MN}+\big(\partial^{M}\xi_{P}-\partial_{P}\xi^{M}\big)*{\cal H}^{PN}+\big(\partial^{N}\xi_{P}-\partial_{P}\xi^{N})*{\cal H}^{MP};
\nn\\
\delta_{\xi} \Phi
&=&-\frac{1}{2}\partial_{M}\xi^{M}+\xi^{M}*\partial_{M}\Phi,     \label{GT}
\eea
where the gauge parameter $\xi_{P}$ is defined by $\xi_{P}\equiv \eta_{PQ} \xi^{Q}$, while the partial derivatives $\partial^{M}$ and $\partial_{M}$ are defined by: $\partial_{M}\equiv\begin{pmatrix}\tilde{\partial}^{j}
&
\partial_{j}
\end{pmatrix}^T$ and $\partial^{M}\equiv\eta^{MQ}\partial_{Q}$.
As already seen for the linearized gauge transformations in Eq. (\ref{gaugetransf}), in the case of non-zero $\lambda$, one chooses:
$\tilde{\xi}_{j}=\xi_{j}$. 
\\

\noindent
It would be very interesting to rewrite $\tilde{S}^{(2)}_{DFT}$ in the generalized metric formulation according to the same steps followed in Ref. \cite{Hohm:2010pp} for $S^{(2)}_{DFT}$ and the cubic action. 
We will leave it as the next possible task, but we make some observations that could be helpful at this aim. 
We will focus therefore on the non-trivial mass terms that have been generated in $\tilde{S}_{add}^{(2)}$.
\\

\noindent
One should observe that, in the context of the generalized metric formulation, the generalized metric by itself \cite{Duff:1989tf,Hohm:2010pp} in Eq. \eqref{HMN} cannot generate the non-trivial massive term involving $\lambda$ since it is constrained to satisfy the equality ${\cal H}^{M P}\eta_{P Q}{\cal H}^{Q N}=\eta^{MN}$ that has to be preserved. 
Such equality implies $\lambda{\cal H}^{MN}{\cal H}_{MN}\sim\lambda$. 
\\

\noindent
Since we cannot generate the non-trivial term with $\lambda$, the suitable gauge transformation for the generalized metric ${\cal H}_{MN}$ should not have an explicit dependence on it. 
Actually, this results to be the case as we are going to discuss.
\\

\noindent
The algebra of the gauge transformations induced in the theory by the double vector $\xi$ can be more deeply analyzed by studying the commutator algebra of the corresponding generalized Lie derivatives. 
Then we first let the corresponding Lie derivative ${\cal L}_{\xi}$  act on a generic scalar field $\tilde{\Phi}$ by calculating ${\cal L}_{\xi}\tilde{\Phi}=\xi^{j}\partial_{j}*\tilde{\Phi}$ to show that the commutator $\lbrack\delta_{\xi_1}, \delta_{\xi_2}\rbrack_*  $, embodying the above star product, defines a closed algebra. 
By direct calculation, one can explicitly get that: $\lbrack{\cal L}_{\xi_1},{\cal L}_{\xi_2}\rbrack_*\tilde{\Phi}=
{\cal L}_{\lbrack\xi_1, \xi_2\rbrack_*}\tilde{\Phi}$,  where $\lbrack\xi_1, \xi_2\rbrack_*\equiv\xi_1^{j}*\partial_{j}\xi_2^{k}-\xi_2^{j}*\partial_{j}\xi_1^{k}$.
This implies that the gauge transformation does not have an explicit dependence on the parameter $\lambda$, and the closure property holds.
\\

\noindent
One can conclude therefore that the gauge transformation of the scalar dilaton field also does not explicitly depend on $\lambda$ since:
\bea
\label{gsd}
\lbrack\delta_{\xi_1}, \delta_{\xi_2}\rbrack_* *\Phi
&=&\frac{1}{2}\partial_{M}\lbrack\xi_1, \xi_2\rbrack_C^{M}-\lbrack\xi_1, \xi_2\rbrack^{M}_C*\partial_{M}\Phi
\nn\\
&=&-\delta_{\lbrack\xi_1, \xi_2\rbrack_C}*\Phi.
\eea
Here the $C$ bracket $[  \cdot \, , \, \cdot  ]_{C} $ is defined by $\left[ \xi_{1}, \xi_{2} \right]_{C}^{M} \equiv  \xi^{N}_{\left[ 1 \right.} \partial_{N} \xi^{M}_{\left. 2 \right]} - \frac{1}{2}  \xi^{P}_{\left[ 1 \right.} \partial^{M} \xi_{\left. 2 \right]P}$  with $\left[ i,j \right] \equiv ij -ji$. 
\\

\noindent
The generalized metric has the same transformation property as the scalar dilaton in Eq. \eqref{gsd} 
$\lbrack\delta_{\xi_1}, \delta_{\xi_2}\rbrack_**{\cal H}^{MN}
=-\delta_{\lbrack\xi_1, \xi_2\rbrack_C}*{\cal H}^{MN}$. 
Hence the closure of the gauge transformations algebra both for the scalar dilaton and the generalized metric holds. 
\\

\noindent
Now we comment on how to probe a generalized metric formulation for the action $\tilde{S}^{(2)}$. 
For a simple extension to this case, we want to retain the O($D$, $D$) notation with its spacetime doubled indices and O($D$, $D$) tensors. 
When ${\cal H}_{MN}$ is promoted to an O($D$, $D$) matrix, constraining it to the condition ${\cal H}\eta{\cal H}=\eta$, one needs to integrate out an auxiliary field $\bar{\lambda}$ from the term $\bar{\lambda}_{MN}({\cal H}\eta{\cal H}-\eta^{-1})^{MN}$ in the action for the generalized metric formulation \cite{Hohm:2010pp}, where the role of the auxiliary field is to reproduce the constraint ${\cal H}\eta{\cal H}=\eta$ through its equation of motion. 
After turning on $\lambda\neq 0$, the constraint ${\cal H}\eta{\cal H}=\eta $ has to be modified by adding to $\eta$ a suitable deformation term  in order to have $\partial_{M}\partial^{M}({\cal H}\eta{\cal H})=-\lambda({\cal H}\eta{\cal H})$, i.e. one has to introduce a suitable deformation necessary for obtaining a consistent relation compatible with the modified constraint. 
Retaining the O($D$, $D$) indices $\big($even losing an O($D$, $D$) element$\big)$ is quite useful because the gauge transformation of the generalized metric only requires the O($D$, $D$) indices without any constraint on the matrix elements.

\section{$d=1$}
\label{sec:4}
\noindent
Now we consider a one-doubled compact direction $d=1$ to solve the modified constraint.
The generic solution of the equation $K_AK^A=\lambda$ 
is given by the $K_A=\begin{pmatrix}
x &
\lambda/(2x)
\end{pmatrix}^T$
for each non-zero constant $x$. 
The momenta $K^B$ and $K^C$ appearing in $B*C$ defined in Eq. \eqref{AB} of course satisfy the same equation. 
Therefore, one has the generic solution for the momenta 
$K_B=\begin{pmatrix}
a&
\lambda/(2a)
\end{pmatrix}^T$ and
$K_C=\begin{pmatrix}
b&
\lambda/(2b)
\end{pmatrix}^T$. 
Then this provides $a/b+b/a=-1$ when $\lambda\neq 0$ from the equation $K_BK_C=-\lambda/2$ . It is easy to show that $a/b$ is an imaginary number, implying no solution with $\lambda\neq 0$ in the case $d=1$. We emphasize that the result is quite general without considering the triple-products. Hence we need to go beyond the $d=1$ case to obtain a non-trivial solution with $\lambda\neq 0$.

\section{Outlook}
\label{sec:5}
\noindent
Following what done in Ref. \cite{Hull:2009mi} in constructing Double Field Theory focused on the string states with $(N_L=1, N_R=1)$, we have constructed the analogous quadratic theory action for the cases $(N_L=2, N_R=0)$ and $(N_L=0, N_R=2)$: these states are indeed massless in the whole $D$-dimensional target-space but massive in the lower dimensional non-compact space. 
Such construction is based on a deformation of the weak constraint through a parameter $\lambda= 2(N_{L}-N_{R})/\alpha^{\prime}$. 
The main appearance of further stringy effects is an extra mass term, given by $\lambda$ itself, for the symmetric two-order metric tensor contained in the levels $(N_L=2, N_R=0)$ and $(N_L=0, N_R=2)$. 
The non-linear gauge transformations with the O($D$, $D$) indices have been defined. 
From there, it turns out that the $\lambda$ parameter only appears in the generalized metric ${\cal H}_{MN}$, but the corresponding gauge transformation does not have an explicit dependence on the parameter, as expected. 
Hence this could simplify a non-linear construction aimed to find a suitable matrix element for ${\cal H}_{MN}$: 
a non-linear extension becomes therefore a possible task.  
We have shown that in the $d=1$ case it is impossible to have a solution of the modified weak constraint that could exhibit stringy effects due to the simultaneous presence of momenta and windings. This implies that it would be interesting to find solutions which are beyond $d=1$ and to explore a non-linear extension of the quadratic theory because all this could give information on further stringy effects in Double Field Theory since the latter does not loose tracks of winding modes.
\\

\noindent
Finally, we would like to make some considerations regarding the energy scale of the effective theory, depending on the mass scale, ${\cal M}^2$. 
For $N_L+N_R=2$, we can obtain ${\cal M}^2=n^2/R^2+m^2R^2/\alpha^{\prime 2}$. 
For $N_L+N_R=3$, the mass scale has the additional $1/\alpha^{\prime}$ term, and it gives ${\cal M}^2=n^2/R^2+m^2R^2/\alpha^{\prime 2}+2/\alpha^{\prime}$. 
When we consider the same $(n, m)$ in both cases, the case of $N_L+N_R=2$ has a lower energy-scale than the one of $N_L+N_R=3$, but it is not enough for the consistent truncation of $N_L+N_R>2$. 
When we consider $(n, m)=(2, 1)$ in $N_L+N_R=2$ and $(n, m)=(1, 3)$ in $N_L+N_R=3$, the mass scale becomes ${\cal M}^2=4/R^2+R^2/\alpha^{\prime 2}$ and ${\cal M}^2=1/R^2+9R^2/\alpha^{\prime 2}+2/\alpha^{\prime}$ respectively, showing that the $N_L+N_R=3$ case cannot be truncated when the compactified radius becomes $R^2\ge\alpha^{\prime}/2$. 
This implies that the consistent effective theory should contain infinite modes, not only the modes from $N_L+N_R=2$. 
We should expect that the different modes can appear simultaneously in the non-linear term. 
Since our study is limited to the quadratic level, it is not necessary to consider such infinite modes from the perspective of the gauge symmetry. 
When we extend the analysis of gauge invariance to the non-linear level, the cancellation of the non-gauge invariance is necessary in order to consider the different constraints simultaneously.

\section*{Acknowledgments}
\noindent
We would like to thank David S. Berman and Jeong-Hyuck Park for their useful discussions.
\\

\noindent
Chen-Te Ma was supported by the Post-Doctoral International Exchange Program and China Postdoctoral Science Foundation, Postdoctoral General Funding: Second Class (Grant No. 2019M652926) and is indebted to Nan-Peng Ma for his encouragement.


\end{document}